\documentclass[ 
    aps,                
    prd,                
    onecolumn,         
    12pts,              
    groupedaddress,     
    superscriptaddress, 
    preprintnumbers,    
    floatfix,
    nofootinbib,        
    titlepage,          
    showkeys,           
    showpacs,           
    ]{revtex4-2}

\usepackage[utf8]{inputenc}
\usepackage{epsfig,amsfonts,amsthm}
\usepackage{epstopdf}
\usepackage{amsfonts,amsmath,amssymb}
\usepackage{latexsym}
\usepackage{bbm}
\usepackage{array}
\usepackage{comment}
\usepackage{hyperref}
\usepackage{color}
\usepackage{pb-diagram}
\usepackage{slashed}
\usepackage{multirow}
\usepackage{cancel}
\usepackage{placeins}                       
\usepackage{diagbox}
\usepackage{fontawesome}

\newcommand{\be}{\begin{equation}}
\newcommand{\ee}{\end{equation}}
\newcommand{\bea}{\begin{eqnarray}}
\newcommand{\eea}{\end{eqnarray}}
\newcommand{\diag}{{\rm{diag}}}

\newcommand{\mean}[1]{\left\langle#1\right\rangle}

\newcommand{\unitv}[1]{\hat{#1}}
\newcommand{\difd}{\,d}

\def\lsim{\mathrel{\raise.3ex\hbox{$<$\kern-.75em\lower1ex\hbox{$\sim$}}}}
\def\gsim{\mathrel{\raise.3ex\hbox{$>$\kern-.75em\lower1ex\hbox{$\sim$}}}}

\begin{document}

\title{Fast numerical evaluation of dark matter direct detection event rates}

\author{Sebastian Sassi}
\email{sebastian.k.sassi@helsinki.fi}
\affiliation{Department of Physics, University of Helsinki, 
                      P.O.Box 64, FI-00014 University of Helsinki, Finland}
\affiliation{Helsinki Institute of Physics, 
                      P.O.Box 64, FI-00014 University of Helsinki, Finland}

\author{Aula Al-Adulrazzaq}
\email{aula.al-adulrazzaq@helsinki.fi}
\affiliation{Department of Physics, University of Helsinki, 
                      P.O.Box 64, FI-00014 University of Helsinki, Finland}
\affiliation{Helsinki Institute of Physics, 
                      P.O.Box 64, FI-00014 University of Helsinki, Finland}

\author{Matti Heikinheimo}
\email{matti.heikinheimo@helsinki.fi}
\affiliation{Department of Physics, University of Helsinki, 
                      P.O.Box 64, FI-00014 University of Helsinki, Finland}
\affiliation{Helsinki Institute of Physics, 
                      P.O.Box 64, FI-00014 University of Helsinki, Finland}

\author{Kimmo Tuominen}
\email{kimmo.i.tuominen@helsinki.fi}
\affiliation{Department of Physics, University of Helsinki, 
                      P.O.Box 64, FI-00014 University of Helsinki, Finland}
\affiliation{Helsinki Institute of Physics, 
                      P.O.Box 64, FI-00014 University of Helsinki, Finland}

\begin{abstract}
\noindent
{The mathematical expression for the dark matter nuclear recoil event rate in a detector consists of a six dimensional integral over the velocity distribution of dark matter in the detector frame, and over the recoil momentum of the nucleus. While a significant part of this integral can be solved in closed form under traditional assumptions of the standard halo model and/or isotropic detector response, analysis of alternate assumptions is conventionally impeded by the inefficiency of multidimensional numerical integration methods. In this work we introduce a novel, fast and efficient algorithm for direct detection event rate computations. This algorithm takes advantage of a closed form expression for the Radon transform of three dimensional Zernike functions, which are used as basis functions for the velocity distribution. We demonstrate an implementation of this algorithm \texttt{ZebraDM} \href{https://github.com/sebsassi/zebradm}{\faGithub}, which is typically faster than an optimized numerical integration approach by a factor between $10^2$ and $10^4$.}
 \end{abstract}
\preprint{HIP-2025-14/TH}
\maketitle

\section{Introduction}
\label{sec:intro}

According to observations, we live in a galaxy filled with dark matter~\cite{Planck:2018vyg}. These observations can be explained within the paradigm of particle dark matter~\cite{Asadi:2022njl}, and direct detection of dark matter (DM) particle(s) is among the most essential goals for contemporary particle physics. The landscape of direct detection experiments is rapidly evolving, and to make use of the vastly expanding experimental data, efficient computation of the expected signal in a given DM model is needed. This calculation is often performed under a collection of simplifying assumptions, namely the standard halo model (SHM) for the DM velocity distribution and isotropic response of the detector to DM scattering events. 

While the SHM is a useful benchmark model of the local velocity distribution, alternatives have been presented~\cite{Evans:2018bqy,Bozorgnia:2019mjk,Smith-Orlik:2023kyl,Reynoso-Cordova:2024xqz,Stanic:2025yze} that likely model the local environment more accurately. At the very least, the impact of the shape of the velocity distribution on dark matter event rates warrants more exploration. The tendency to adopt SHM in analyses of direct detection is at least in part due to the closed form solution of the event rate it enables, whereas more general distributions typically require numerical integration, which significantly slows down the analysis, reducing the scope of what can be done with limited computational resources.

Secondly, in the study of sub-GeV dark matter, the impact of potential anisotropies of a detector's response on the dark matter event rate has been recognized~\cite{Kadribasic:2017obi,Griffin:2018bjn,Trickle:2019nya,Heikinheimo:2019lwg,Coskuner:2021qxo,Campbell-Deem:2022fqm,Stratman:2024sng}. Due to the rotation of the Earth, these anisotropies induce a daily modulation on top of the well-known annual modulation of the direct detection signal. However, the same causes that generate the annual modulation cause the features of the daily modulation signal to vary throughout the year~\cite{Sassi:2021umf,Heikinheimo:2023dtz}. Therefore, proper analysis of the signal modulation requires sampling the event rate multiple times a day, throughout the entire year, requiring on the order of $10^4$ time points. Since the anisotropic response always needs to be integrated numerically, such an analysis easily becomes prohibitively expensive.

These facts indicate an urgent need for more efficient methods of evaluating the dark matter event rate for general velocity distributions beyond the SHM, in combination with anisotropic models of detector response. Besides the above concrete examples, in general it is undeniable that in the history of physics the invention of better numerical methods has always enabled a broader scope of study, and thus led to new and unexpected discoveries. Therefore in the study of dark matter direct detection, too, we should seek and adopt better, more efficient numerical methods.

The bottleneck in the computation of dark matter scattering rates is the dimensionality of the relevant integrals. In the case of dark matter scattering on atomic nuclei, for example, the total event rate at a given time is in the simplest case given by a six dimensional integral. This integral consists of a three-dimensional integral over the dark matter velocity, a two-dimensional integral over the recoil direction of the target nucleus, and a one-dimensional integral over the recoil energy. Although the integrand contains a delta-function due to kinematic constraints, which enables reduction of the dimensionality by one, a five dimensional integral remains to be evaluated numerically in the general case. Direct numerical integration of such a high dimensional integral is prohibitively expensive. Furthermore, in other cases the dimensionality may be even higher, for example, if the target particle cannot be assumed to be initially at rest, as is the case in electron scattering.

To address these challenges, in this paper we propose a novel algorithm based on expansion of the velocity distribution in a basis of so-called three dimensional Zernike functions, which are polynomials of the Cartesian coordinates and are orthogonal on the unit ball~\cite{Canterakis1997}. We take advantage of the fact that as a consequence of the kinematic constraints, the velocity integral is equivalent to a three-dimensional Radon transform, which can be evaluated in closed form for the Zernike functions. This leads to a highly efficient algorithm for the evaluation of angle-integrated Radon transforms---and therefore dark matter event rates---which is orders of magnitude faster than plain numerical integration.

The general idea of expansion in a suitable basis of functions to solve the event rate integrals is not new, and has been discussed in various context in the literature. For example, an expansion of the velocity distribution in a basis built from spherical harmonics and spherical Bessel functions was discussed in~\cite{Lee:2014cpa}, whereas more recently in~\cite{Lillard:2023cyy, Lillard:2023nyj} a spherical analog of Haar wavelets was used for the radial functions.

The computational advantages of a basis expansion are evident: if the relevant scattering rate integrals have closed form solutions for the basis functions, then the integrals can be traded for sums, which lend themselves to more efficient numerical evaluation for a given target accuracy than the equivalent integrals. However, the choice of basis significantly impacts the practicality of this. For the angular part of the basis functions, spherical harmonics are the natural choice, but for the radial part there is no single obvious choice.

The choice of spherical Bessel functions is natural for a function defined on $\mathbb{R}^3$, and the Radon transform of the resulting basis functions has closed form solutions. However, the radial part is indexed by a continuous variable and therefore a finite truncation cannot be taken without restricting the domain. Furthermore, the resulting expressions for the Radon transform are complicated and difficult to manipulate, which poses a challenge for their practical usefulness for the task at hand.

On the other hand, while the Haar-like wavelets may appear attractive due to their simplicity, being piecewise constant functions, their discontinuity results in poor convergence for the expansion of any smooth function. Although they may be straightforward to deal with, the large number of expansion coefficients needed for an accurate expansion due to the poor convergence behavior is suboptimal.

In this paper, we propose to use the three diemensional Zernike functions as a basis, because they possess several desirable properties: First, as polynomials of the Cartesian coordinates, they naturally approximate smooth functions with good convergence properties, as we will demonstrate. Second, they can be computed rapidly to high orders using recursion formulae. Third, their Radon transforms have known polynomial expressions, which are straightforward to manipulate~\cite{Louis1984, Janssen2015}.

An apparent limitation of Zernike functions is the restriction of their orthogonality to the unit ball. However, this is not a real issue for two reasons: First, for many common dark matter velocity distributions the Milky Way escape velocity $v_\text{esc}$ presents a natural cutoff, which means the integral is evaluated over a ball of some radius. Second, even distributions with no explicit cutoff generally fall off exponentially at high velocities. In such a situation one can always find some cutoff velocity beyond which the contribution to the distribution is negligible. In fact, in practice one typically has to take some cutoff, as in the case of the spherical Bessel functions mentioned above, in order to have an expansion with a finite number of coefficients. Reduction from a ball of arbitrary radius to a unit ball is then just a matter of rescaling the coordinates.

Although this analysis focuses on the numerical aspects of event rate computations, it is worth pointing out that a basis function expansion of the velocity distribution has other uses. For instance, it can be used for the reconstruction problem: given some data on observed event rates, the basis expansion coefficients can be used to fit the velocity distribution on the data in a general, model agnostic way. This has been a motivation in many prior uses of basis expansions~\cite{Lee:2014cpa, Alves:2012ay, Kavanagh:2013eya, Peter:2013aha}. While it is plausible that the Zernike basis suggested here could also be suitable for such purposes, such an analysis is beyond the scope of this study.

The structure of this paper is as follows. Sec.~\ref{sec:general} starts with a brief review of the relevant formulae for dark matter event rate computations and describes the problems surrounding their numerical evaluation. In Sec.~\ref{sec:zernike} the three dimensional Zernike functions and their properties are introduced and a fast algorithm for transforming an arbitrary function on the unit ball into a basis of three dimensional Zernike functions is described. This algorithm is essentially an extension of a well known algorithm for spherical harmonic transforms. Next, in Sec.~\ref{sec:radon} the properties of the Zernike functions established in the previous section are applied to develop a method for efficient evaluation of the Radon transform of a shifted distribution inside the unit ball. Finally, Sec.~\ref{sec:benchmark} contains an analysis of the accuracy, convergence properties, and performance of the methods developed in the previous sections for a number of test cases. Some of the detailed formulas are provided in the Appendix~\ref{app:trradon}.

We provide a reference implementation of our method in two C$++$ libraries. The library \texttt{ZebraDM} \cite{zebradm} implements the main Zernike-based Radon transform algorithm presented in this paper. It also contains an implementation of a conventional numerical integration algorithm for comparison, as well as the code for the benchmark results presented in this paper. The companion library \texttt{zest} \cite{zest} implements data structures and algorithms for dealing with Zernike and spherical harmonic basis expansions, including forward and backward transforms, as well as rotations of the basis coefficients.

\section{Dark matter event rates}
\label{sec:general}

To motivate the methods developed in this paper, we provide a brief outline of the equations relevant to dark matter direct detection. The methods presented in the subsequent sections of this paper are broadly applicable to any scattering problem where the dark matter has a well-defined velocity distribution, and energy conservation can be written in the form
\begin{equation}
\vec{v}\cdot\unitv{q}-w(\vec{q},\ldots)=0.
\label{eq:kinematic-constraint}
\end{equation}
In the above equation $\vec{q}$ is the momentum transfer from dark matter to the target, and $w(\vec{q},\ldots)$ may depend on the momentum transfer $\vec{q}$ and other parameters denoted by the dots, but, crucially, does not depend on $\vec{v}$. For example, as described in~\cite{Trickle:2019nya}, spin-independent dark matter scattering producing nuclear recoils, electron transitions, and single-phonon excitations can be described in a unified framework which has the desired properties.

In this work we assume that $w$ does not depend on the direction of the momentum transfer. Note that this does not imply any loss of generality, because it is always possible to write
\begin{equation}
\delta(\vec{v}\cdot\unitv{q}-w(\vec{q},\ldots))=\int\delta(\vec{v}\cdot\unitv{q}-W)\delta(W-w(\vec{q},\ldots))\difd W.
\label{eq:nop}
\end{equation}
An equivalent transformation has been employed in~\cite{Catena:2021qsr} in the context of electron scattering. Note that although this introduces an extra integration parameter, it is via the introduction of a delta-function and therefore does not contribute to the numerical load. It effectively factors the complex $\vec{q}$-dependence of $w$ into evaluation of the detector response.

For simplicity we will focus on the case of dark matter scattering on nuclear targets.
The differential scattering rate in a general dark matter nuclear scattering process is~\cite{Gondolo:2002np}
\begin{equation}
\frac{d^2R_S}{dE_{\vec q}d\Omega}=\frac{1}{64\pi^2}\frac{\rho_0}{m_\text{DM}^3m_\text{N}^2}\int\mean{|\mathcal{M}|^2}\delta(\vec{v}\cdot\unitv{q}-v_\text{min})f(\vec{v})\difd^3 v.
\label{eq:dmrate1}
\end{equation}
Here $E_{\vec q}$ is the nuclear recoil energy, $\rho_0$ is the local dark matter density, $m_\text{DM}$ is the mass of the dark matter particle, and $m_\text{N}$ is the mass of the target nucleus. The factor $\mathcal{M}$ is the scattering amplitude, and $f(\vec{v})$ is the dark matter velocity distribution. The parameter $v_\text{min}=w(\vec{q})$ is given by
\begin{equation}
v_\text{min}=\frac{q}{2\mu_\text{DM,N}},
\end{equation}
where $\mu_\text{DM,N}$ is the reduced mass of the DM--nucleus system.

The form of the scattering amplitude 
depends on the details of the interaction of dark matter with the target nucleus. However, the scattering of dark matter on nuclei can be described in the framework of a nonrelativistic effective theory~\cite{Fitzpatrick:2012ix,Anand:2013yka}. We will not describe details of the theory here, but will only point out a result relevant for this work, which is that for nearly all possible interactions we may write
\begin{equation}
\mean{|\mathcal{M}|^2}=F(q^2) + F_\perp(q^2)|\vec{v}_\perp|^2,
\label{eq:amplitudepoly}
\end{equation}
where $\vec{v}_\perp$ is the component of the dark matter velocity perpendicular to the recoil direction $\unitv{q}$, and $F(q^2)$ and $F_\perp(q^2)$ capture the details of the interactions. An analogous framework for electron scattering has been described in~\cite{Catena:2019gfa, Catena:2021qsr}. The definition of $\vec{v}_\perp$ in this framework is slightly different, and is not perpendicular to $\vec{q}$ anymore. Therefore the ionization amplitude also gains terms proportional to $(\vec{v}_\perp\cdot\vec{q})^2$. We do not address the issue of such terms in the present work, but note that they can likely be dealt with in a similar manner as $|\vec{v}_\perp|^2$ terms.

Given the parameterization in Eq.~\eqref{eq:amplitudepoly}, we define the Radon transform integrals
\begin{align}
\label{eq:radon}
\mathcal{R}[f](\unitv{q},v_\text{min})&=\int\delta(\vec{v}\cdot\unitv{q}-v_\text{min})f(\vec{v})\difd^3 v,\\
\label{eq:transverse-radon}
\mathcal{R}_\perp[f](\unitv{q},v_\text{min})&=\int(\vec{v}_\perp)^2\delta(\vec{v}\cdot\unitv{q}-v_\text{min})f(\vec{v})\difd^3 v.
\end{align}
We refer to these as the (standard) Radon transform, and the transverse Radon transform, respectively. Then the integral in Eq.~\eqref{eq:dmrate1} reduces to
\begin{equation}
\frac{d^2R_S}{dE_{\vec q}d\Omega}=\frac{1}{64\pi^2}\frac{\rho_0}{m_\text{DM}^3m_\text{N}^2}(F(q^2)\mathcal{R}[f](\unitv{q},v_\text{min})+F_\perp(q^2)\mathcal{R}_\perp[f](\unitv{q},v_\text{min})).
\label{eq:dmrate}
\end{equation}

The scattering rate itself is generally not observable in an experiment, because not all scattering events lead to observable signals. We can write the rate of events as a function of the observed energy $E$ generally as
\begin{equation}
\frac{d^2R_E}{dEd\Omega}=\int S(E,E_{\vec{q}},\unitv{q})\frac{d^2R_S}{dE_{\vec{q}}d\Omega}\difd E_{\vec{q}},
\label{eq:event-rate}
\end{equation}
where the response function $S(E,E_{\vec{q}},\unitv{q})$ is the probability density for a nuclear recoil with recoil direction $\unitv{q}$ and recoil energy $E_{\vec{q}}$ to be observed with measured energy $E$. This function captures the material properties of the target as well as smearing due to finite energy resolution etc.
The exact definition of the response function is not relevant here and depends on the context. This method works equally for any response function which can be written as a function of the momentum transfer $\vec{q}$ or, equally, as a function of the energy transfer $E_{\vec{q}}$ and the direction $\unitv{q}$.

Typically, the velocity distribution $f(\vec{v})$ is taken to be that of the standard halo model (SHM), which is given by a simple isotropic Gaussian function
\begin{equation}
f_\text{I}(\vec{v})=\frac{1}{N}e^{-v^2/v_0^2}\Theta(v_\text{esc}-v).
\label{eq:isodist}
\end{equation}
Here $N$ is a normalization constant and the velocity dispersion is given by $\sigma_v=v_0/\sqrt{2}$, where $v_0$ is the local circular speed of the Milky Way at the location of the solar system, and the escape velocity of the Milky Way $v_\text{esc}$ acts as an upper limit for dark matter velocities.

For the SHM distribution the integrals in Eqs.~\eqref{eq:radon} and~\eqref{eq:transverse-radon} are straightforward to evaluate analytically. However, the SHM is likely to be a simplification of the true velocity distribution. For example, recent studies suggest at least the presence of an anisotropic component in addition to the isotropic one, motivated by observations from the Gaia satellite, which gives rise to the SHM$^{++}$ model presented in~\cite{Evans:2018bqy}. The form of this 
velocity distribution is
\be
f_{{\rm SHM}^{++}}(\vec v) = (1-\eta)f_\text{I}(\vec v) + \eta f_\text{A}(\vec v),
\label{eq:SHM++}
\ee
where $f_\text{I}$ is the isotropic component \eqref{eq:isodist}, $f_\text{A}$ is the anisotropic component and $0 \leq \eta \leq 1$ is the anisotropy fraction. The velocity distribution for the anisotropic component is
\be
f_\text{A}(\vec v)=\frac{1}{N}e^{-\frac{1}{2}\vec{v}^T\Sigma^{-1}\vec{v}}\Theta(v_{\rm{esc}}-v),
\label{eq:anisodist}
\ee
where $\Sigma=\diag(\sigma_r^2,\sigma_\theta^2,\sigma_\phi^2)$ in the galactic frame, with
\be
\sigma_r^2=\frac{3v_0^2}{2(3-2\beta)},\quad\sigma_\theta^2=\sigma_\phi^2=\frac{3v_0^2(1-\beta)}{2(3-2\beta)}.
\label{eq:dispersion}
\ee
From~\cite{Evans:2018bqy}, the value of the circular rotation speed in SHM$^{++}$ is $v_0=233$ km s$^{-1}$, the escape velocity is $v_{\rm{esc}}=528$ km s$^{-1}$ and the anisotropy parameter  $\beta$ is 0.9. Another description of the velocity distribution motivated by the Gaia observations is the self-consistent anisotropic halo model (scAHM) given in~\cite{Stanic:2025yze}. This model does not have a simple analytical form in terms of the local dark matter velocity.

Therefore, there exists a clear motivation to examine non-SHM velocity distributions in the context of direct detection of dark matter. However, these alternative distributions typically do not have analytic solutions for the velocity integrals, and are therefore significantly more expensive to compute event rates with. Studies and adoption of such alternative velocity distributions in the context of direct detection experiments would therefore benefit greatly from more efficient computational methods for evaluation of the event rate integrals.

The scattering of dark matter particles is not observed in the rest frame of the velocity distribution, which is the galactic frame, but usually in an Earth-based laboratory. Therefore, in Eqs.~\eqref{eq:radon},~\eqref{eq:transverse-radon}, and~\eqref{eq:dmrate} the velocity $\vec{v}$ is replaced with $\vec{v}+\vec{v}_\text{lab}$, where $\vec{v}_\text{lab}$ is the velocity of the laboratory in the galactic frame. It can be expressed in terms of its various contributions as $\vec{v}_\text{lab}=\vec{v}_\text{circ}+\vec{v}_\text{pec}+\vec{v}_\text{earth}+\vec{v}_\text{rot}$~\cite{Mayet:2016zxu}. Here $\vec{v}_\text{circ}$ is the circular velocity of the solar system in the galactic plane, $\vec{v}_\text{pec}$ is the peculiar velocity, $\vec{v}_\text{earth}$ is the velocity of the Earth around the sun, and $\vec{v}_\text{rot}$ is the rotational velocity of the point on Earth's surface. The variation in the magnitude of $\vec{v}_\text{lab}$ due to the changing direction of $\vec{v}_\text{earth}$ results in the well known annual modulation of the dark matter event rate.

Due to the motion of the laboratory frame relative to the dark matter, the directional scattering rate has an anisotropy, which commonly appears as a maximum in the direction of $\vec{v}_\text{lab}$. If the response in Eq.~\eqref{eq:event-rate} is anisotropic, this can induce a daily modulation in the angle-integrated event rate, due to the rotation of the Earth. However, because $\vec{v}_\text{lab}$ changes direction throughout the year, the phase and amplitude of the daily modulation can vary in a complex manner throughout the year~\cite{Sassi:2021umf, Heikinheimo:2023dtz}. As a result, accurate analysis including both annual and daily modulation requires evaluation of the event rate at around $10^4$ time points. Such analyses present an urgent need for faster means of evaluating the event rate integrals.

In a typical direct detection analysis, one is interested in particular in the angle-integrated event rate
\begin{equation}
\frac{dR_E}{dE}=\int S(E,E_{\vec{q}},\unitv{q})\frac{d^2R_S}{dE_{\vec{q}}d\Omega}\difd\Omega\difd E_{\vec{q}}.
\end{equation}
In the case of a general anisotropic response, and an anisotropic velocity distribution such as that in Eq.~\eqref{eq:SHM++}, this amounts to evaluation of a six-dimensional integral. The delta-function in the Radon transforms (see Eqs.~\eqref{eq:radon} and~\eqref{eq:transverse-radon}) allows trivial closed form evaluation of one of these integrals, leaving a five-dimensional integral to be evaluated numerically. In this work we develop an algorithm which enables fast evaluation of four-dimensional integrals of the form
\begin{equation}
\frac{d^2R_E}{dEdE_{\vec{q}}}=\int S(E,E_{\vec{q}},\unitv{q})\frac{d^2R_S}{dE_{\vec{q}}d\Omega}\difd\Omega.
\label{eq:double-energy-rate}
\end{equation}
Although an extension of our method to evaluating the energy integral is straightforward, we do not consider such an extension here. There are circumstances where a separate evaluation of~\eqref{eq:double-energy-rate} is advantageous, for example, because the integrand can be made independent of the exact form of dark matter interactions. We therefore deem it more general to present a solution of~\eqref{eq:double-energy-rate}, since the ideal way of dealing with the energy integral depends on the specifics of the problem at hand.

\section{Zernike functions}
\label{sec:zernike}

The three dimensional Zernike functions $Z_{nlm}^\alpha(\rho,\theta,\varphi)$ are a family of functions on $\mathbb{R}^3$ and orthogonal within the unit ball $B=\{\vec{x}\in\mathbb{R}^3 : |\vec{x}|\leq1\}$.
They are three dimensional generalization of the better known two dimensonal Zernike functions, which find applications in various fields~\cite{NiuTiang2022}. Although in the literature ``Zernike polynomials'' and ``Zernike functions'' commonly refer to the two dimensional case, in this paper we use these terms to refer to the three dimensional functions for brevity. Explicitly, they can be defined as
\begin{equation}
Z_{nlm}^\alpha(\rho,\theta,\varphi)=R_n^{l,\alpha}(\rho)Y_{lm}(\theta,\varphi),
\label{eq:zernike}
\end{equation}
where the unnormalized radial Zernike functions are
\begin{equation}
R_n^{l,\alpha}(\rho)=\rho^l(1-\rho^2)^\alpha P_{(n-l)/2}^{(\alpha,l+1/2)}(2\rho^2-1),
\label{eq:radialzernike}
\end{equation}
with $P^{(\alpha,\beta)}_k(x)$ Jacobi polynomials and $Y_{lm}(\theta,\varphi)$ real spherical harmonics~\cite{Canterakis1997, Mathar:2008qe, Janssen2015}. The parameter $\alpha\geq0$ is a free parameter. The indices here have the hierarchy $0\leq|m|\leq l\leq n$ with the added restriction that $(n-l)/2$ must be an integer, and hence $n-l$ must be even, which in turn implies that $n$ and $l$ must have the same parity. The unnormalized radial Zernike functions for given $\alpha$ and $l$ are orthogonal with
\begin{equation}
\int_0^1 R_n^{l,\alpha}(\rho)R_{n'}^{l,\alpha}(\rho)\frac{\rho^2\difd\rho}{(1-\rho^2)^\alpha}=N_{nl}^\alpha\delta_{nn'},
\end{equation}
where the normalization factor is
\begin{equation}
N_{nl}^\alpha=\frac{1}{2(n+\alpha+3/2)}\frac{((n-l)/2+1)_\alpha}{((n-l)/2+l+3/2)_\alpha},
\end{equation}
where $(x)_\alpha$ denotes the rising factorial, $(x)_\alpha=x(x+1)\cdots(x+\alpha-1)$. It follows from this and the orthogonality of the spherical harmonics that, for a given value of $\alpha$, the Zernike functions are orthogonal over the unit ball with
\begin{equation}
\int_BZ_{nlm}^\alpha(\vec{x})Z_{n'l'm'}^\alpha(\vec{x})\frac{d^3x}{(1-|\vec{x}|^2)^\alpha}=N_{nlm}^\alpha\delta_{nn'}\delta_{ll'}\delta_{mm'},
\end{equation}
where $B$ denotes the unit ball, and $N_{nlm}^\alpha$ is a normalization factor depending on the normalization convention of both the radial Zernike functions, and of the spherical harmonics.

It is worth noting that one can alternatively construct a collection of complex Zernike functions $Z_{nl}^m$ using the complex spherical harmonics $Y_l^m$. However, since we are dealing with real functions in this analysis, the choice of real basis functions is a more natural one. Because the real spherical harmonics are given by a simple linear combination of the complex spherical harmonics and vice versa, the results shown here for the real functions should generally be applicable to the complex functions up to a linear transform.

The properties of the radial Zernike functions are straightforward to derive using properties of the Jacobi polynomials. For example, the following recursion relation can be proven for the radial Zernike functions~\cite{Deng2020},
\begin{equation}
R_n^l(\rho)=(K_1^{nl}\rho^2+K_2^{nl})R_{n-2}^l(\rho)+K_3^{nl}R_{n-4}^l(\rho),
\label{eq:radialzernike}
\end{equation}
with the coefficients
\begin{align}
K_1^{nl}&=\frac{(2n-1)(2n+1)}{(n-l)(n+l+1)},\\
K_2^{nl}&=-\frac{(2n-1)((2l+1)^2+(2n+1)(2n-3))}{2(n-l)(n+l+1)(2n-3)},\\
K_3^{nl}&=-\frac{(n-l-2)(n+l-1)(2n+1)}{(n-l)(n+l+1)(2n-3)}.
\end{align}
This recursion relation is valid for $l\leq n-4$. For the remaining indices we can see from the definition that
\begin{align}
R_n^n(\rho)&=\rho^n,\\
R_n^{n-2}&=\frac{2n+1}{2}\rho^n-\frac{2n-1}{2}\rho^{n-2}.
\end{align}

The main attractive property of the Zernike functions for our purposes is that, given a Radon transform defined as
\begin{equation}
\mathcal{R}[f](w,\unitv{n})=\int_B\delta(w-\vec{x}\cdot\unitv{n})f(\vec{x})\difd^3x,
\label{eq:radon-2}
\end{equation}
the Radon transform of the functions $Z_{nlm}^\alpha$ is given by~\cite{Louis1984, Janssen2015}
\begin{equation}
\mathcal{R}[Z_{nlm}^\alpha](w,\unitv{n})=\frac{1}{c^\alpha_{nl}}(1-w^2)^{\alpha+1}C_n^{\alpha+3/2}(w)Y_{lm}(\unitv{n}),
\label{eq:radzernike}
\end{equation}
where
\begin{equation}
c_{nl}^\alpha=\frac{2^{-2(1+\alpha)}}{\sqrt{\pi}((n-l+2)/2)_\alpha}\frac{\Gamma(n+2\alpha+3)}{\Gamma(n+1)\Gamma(\alpha+3/2)},
\end{equation}
and $C_n^{\alpha+3/2}(x)$ are Gegenbauer polynomials. This result is notable, because 
if we have a function defined as a linear combination of Zernike functions,
\begin{equation}
f(\vec{x})=\sum_{n=0}^L\sum_{l=0}^n\sum_{|m|\leq l}f_{nlm}Z^\alpha_{nlm}(\vec{x}),
\end{equation}
by linearity the Radon transform of $f(\vec{x})$ is
\begin{equation}
\mathcal{R}[f](w,\unitv{n})=\sum_{n=0}^L\sum_{l=0}^n\sum_{|m|\leq l}f_{nlm}\mathcal{R}[Z_{nlm}^\alpha](w,\unitv{n}).
\end{equation}

The preceding results hold for general $\alpha$, but due to the presence of the term $(1-w^2)^{\alpha+1}$ Eq. \eqref{eq:radzernike}, the choice of $\alpha=0$ turns out to be simplest case to deal with. From this point on we will restrict to that case, denoting $Z_{nlm}\equiv Z_{nlm}^0$.

A quantity of practical importance for the numerical evaluation of the Zernike expansion of a function is the number of coefficients needed for an expansion. If we consider an expansion to order $L$, such that we have a pair of coefficients (for $+m$ and $-m$) for each $0\leq|m|\leq l\leq n\leq L$, the number of pairs in the expansion is given by the sequence $1,3,7,13,22,34,50,\ldots$
Therefore, the total number of coefficients up to order $L$ is\footnote{The sequence is identified as A002623 in The On-Line Encyclopedia of Integer Sequences (OEIS)~\cite{oeis}.}
\begin{equation}
N(L)=2\left\lfloor\frac{(L+2)(L+4)(2L+3)}{24}\right\rfloor,
\label{eq:num-coeff}
\end{equation}
where $\lfloor \cdot \rfloor$ denotes the floor function. 
Note that, since for $m=0$ there is only one coefficient, this overcounts the total number of coefficients by a small margin. However, pairwise storage of the coefficients turns out to be computationally advantageous. In any case, we see from here that the number of cofficients grows as $N(L)\approx L^3/6$ for large $L$.

The Zernike functions and their linear combinations benefit from sharing the usual properties of spherical harmonics. For example, under reflection $Z_{nlm}(-\vec{x})=(-1)^lZ_{nlm}(\vec{x})$, and the Zernike functions can be rotated using Wigner $D$-matrices,
\begin{equation}
Z_{nlm}(R\vec{x})=\sum_{|m'|\leq l}\mathcal{R}_{mm'}^l(R)Z_{nlm'}(\vec{x}),
\end{equation}
where $\mathcal{R}_{mm'}^l(R)$ denotes the real spherical harmonic equivalent of the complex Wigner $D$-matrix for a rotation $R$.

Existing methods for transformation of functions defined on the sphere to the spherical harmonic basis can be extended to transformations of functions in the unit ball to the Zernike basis. One such relatively straightforward and efficient method, used in various spherical harmonic codes (see, e.g.,~\cite{Schaeffer2013, Wieczorek2018}), is based on evaluating the function $f(\vec{x})$ on a spherical grid with equispaced longitudes and Gauss--Legendre quadrature nodes for the latitudes. This method has a straightforward extension to the unit ball if the radii are also chosen by mapping the Gauss--Legendre nodes to the interval $[0,1]$. This method allows us to replace the integrals
\begin{equation}
f_{nlm}=\int_0^1\int_{-1}^1\int_0^{2\pi}f(\rho,\theta,\varphi)R_n^l(\rho)P_l^m(\cos\theta)\begin{Bmatrix}\cos (m\varphi)\\\sin(|m|\varphi)\end{Bmatrix}\difd\varphi\difd\cos\theta\rho^2\difd\rho
\end{equation}
with the sums
\begin{equation}
f_{nlm}\approx\sum_{ijk}w_i\rho_i^2R_n^l(\rho_i)w_jP_l^m(\cos\theta_j)\begin{Bmatrix}\cos (m\varphi)\\\sin(|m|\varphi)\end{Bmatrix}f(\rho_i,\theta_j,\varphi_k),
\label{eq:glsum}
\end{equation}
where $w_i$ are Gauss--Legendre quadrature nodes. An appealing property of this method is that if $f(\vec{x})$ is a linear combination of Zernike functions up to order $L$, then the sum is exact for a grid of size\footnote{The radial direction requires an additional node compared to the latitudinal direction due to the $\rho^2$ factor coming from the integration measure.} $(L+2)(L+1)(2L+1)$. In fact, the truncated expansion of $f(\vec{x})$ computed this way is an interpolating polynomial on these nodes and as such benefits from the good convergence properties of polynomial interpolation. This property is a result of the fact that $N$ Gauss--Legendre nodes are sufficient to integrate a polynomial of order $2N - 1$ exactly.

Naively, the summation in \eqref{eq:glsum} looks like it would involve $O(L^6)$ operations---$O(L^3)$ operations for each triple $(n,l,m)$. However, more careful observation reveals it can be split into three steps with at most $O(L^4)$ operations each. First, the sums
\begin{equation}
f_m(\rho_i,\theta_j)=\sum_k\begin{Bmatrix}\cos (m\varphi)\\\sin(|m|\varphi)\end{Bmatrix}f(\rho_i,\theta_j,\varphi_k)
\label{eq:stepfm}
\end{equation}
can be performed using a fast Fourier transform with a total of $O(L^3\log L)$ operations. Next, the sums
\begin{equation}
f_{lm}(\rho_i)=\sum_kw_jP_l^m(\cos\theta_j)f_m(\rho_i,\theta_j)
\label{eq:stepflm}
\end{equation}
are performed with $O(L^4)$ operations, and finally the sums
\begin{equation}
f_{nlm}=\sum_{ijk}w_i\rho_i^2R_n^l(\rho_i)f_{lm}(\rho_i)
\label{eq:stepfnlm}
\end{equation}
take another $O(L^4)$ operations. Steps~\eqref{eq:stepfm} and~\eqref{eq:stepflm} together essentially constitute a collection of spherical harmonic transforms for each radial grid point $\rho_i$.

\section{Radon transform of a shifted distribution}
\label{sec:radon}

In the theoretical description of dark matter scattering on Earth-based targets, kinematic constraints on the scattering momenta result in the double differential scattering rate being proportional to a Radon transform, which can be expressed in a form
\begin{equation}
\mathcal{R}[f](w,\unitv{n})=\int\delta(w-\vec{x}\cdot\unitv{n})f(\vec{x} + \vec{x}_0)\difd^3x.
\label{eq:shiftradon}
\end{equation}
Here $f(\vec{x})$ would be the local velocity distribution of dark matter in a frame moving at velocity $\vec{x}_0$ relative to the avarage motion of the dark matter. The factor $w$ is that of Eq.~\eqref{eq:kinematic-constraint}. In general, $w$ could also be a function of $\unitv{n}$, however, using Eq.~\eqref{eq:nop} more general integrals can be reduced to working with a constant $w$.

In many models of the dark matter halo the distribution is cut off at the local escape velocity. In appropriately scaled units this means that $|\vec{x}+\vec{x}_0|\leq1$ (namely, by defining $\vec{x}=\vec{v}/v_\text{esc}$). This fact renders the Zernike basis an ideal choice of basis for velocity distributions. However, this is not a limitation on the usefulness of the Zernike expansion in the case when there is no natural cutoff. Practically all physically meaningful distributions decay exponentially at high velocities, so even for distributions that extend beyond $v_\text{esc}$ some finite cutoff can always be found beyond which the contributions to the distribution can be neglected. For example, a narrow stream with velocities of the order of $v_\text{esc}$ and dispersion around $0.1v_\text{esc}$ can be cut off at $1.5v_\text{esc}$ without meaningful loss of accuracy.

In any case, for evaluation of the integral~\eqref{eq:shiftradon}, it is not desirable to perform the Zernike expansion in these coordinates, because the ball in which the distribution has support is centered on $\vec{x}_0$. Therefore we transform to new integration coordinates via $\vec{x}+\vec{x}_0\rightarrow\vec{x}$ such that the integral becomes
\begin{equation}
\mathcal{R}[f](w,\unitv{n})=\int_B\delta(w+\vec{x}_0\cdot\unitv{n}-\vec{x}\cdot\unitv{n})f(\vec{x})\difd^3x,
\end{equation}
where we have also accounted for the cutoff so that we are in the situation of Eq.~\eqref{eq:radon-2}. If we now substitute $f(\vec{v})$ with its truncated Zernike expansion, choosing the Zernike functions with $\alpha=0$, we obtain
\begin{equation}
\mathcal{R}[f](w,\unitv{n})=2\pi\sum_{n=0}^L\sum_{l=0}^n\sum_{|m|\leq l}\frac{f_{nlm}}{(n+1)(n+2)}(1-(w+\vec{x}_0\cdot\unitv{n})^2)C_n^{3/2}(w+\vec{x}_0\cdot\unitv{n})Y_{lm}(\unitv{n}).
\label{eq:radonfseries}
\end{equation}
If we only wish to evaluate the Radon transform for some triple $(w,\vec{x}_0,\unitv{n})$, this formula is in principle sufficient. However, in analysis of dark matter scattering we often need to integrate over the scattering directions $\unitv{n}$, and therefore it would be beneficial to be able to directly compute the angle integrated Radon transforms. Eq.~\eqref{eq:radonfseries} can be simplified by using a recursion relation of the Gegenbauer polynomials,
\begin{equation}
2\lambda (1-x^2)C^{\lambda+1}_n(x)=\frac{(n+2\lambda)(n+2\lambda+1)C^\lambda_{n}(x)-(n+1)(n+2)C^\lambda_{n+2}}{2(n+\lambda+1)}.
\label{eq:general-gegursion}
\end{equation}
This can be derived from the two conventional recursion relations
\begin{align}
(n+1)C^{\lambda}_{n+1}(x)&=2(n+\lambda)xC^{\lambda}_n(x)-(n+2\lambda-1)C^{\lambda}_{n-1}(x),\\
2\lambda(1-x^2)C^{\lambda+1}_{n-1}(x)&=(2\lambda+n-1)C^{\lambda}_{n-1}(x)-nxC^{\lambda}_n(x).
\end{align}
For the special case of $\lambda=3/2$ the relation~\eqref{eq:general-gegursion} reduces to
\begin{equation}
(1-x^2)C^{3/2}_n(x)=(n+1)(n+2)\frac{C^{1/2}_{n}(x)-C^{1/2}_{n+2}(x)}{2n+3}.
\label{eq:gegursion}
\end{equation}
We may further note that the Gegenbauer polynomials at $\lambda=1/2$ are just the Legendre polynomials, $C^{1/2}_{n}(x)=P_n(x)$. Therefore, Eq.~\eqref{eq:gegursion} allows us to write Eq.~\eqref{eq:radonfseries} as
\begin{equation}
\mathcal{R}[f](w,\unitv{n})=2\pi\sum_{n=0}^{L+2}\sum_{l=0}^n\sum_{|m|\leq l}f'_{nlm}P_n(w+\vec{x}_0\cdot\unitv{n})Y_{lm}(\unitv{n}),
\label{eq:radonfprime}
\end{equation}
where
\begin{equation}
f'_{nlm}=\frac{1}{2n+3}f_{nlm}-\frac{1}{2n-1}f_{n-2,lm}.
\label{eq:radon-fnlm}
\end{equation}
Here the second term is neglected for $n=0,1$ and $f_{nlm}=0$ for $n>L$ and $l>n$.

We can see that the integration of~\eqref{eq:radonfprime} over the directions $\unitv{n}$, reduces to evaluation of the integrals
\begin{equation}
\int\displaylimits_{|w+\vec{x}_0\cdot\unitv{n}|\leq1}\hspace{-1.7em}P_n(w+\vec{x}_0\cdot\unitv{n})Y_{lm}(\unitv{n})\difd\Omega.
\end{equation}
Note that the integration limits follow from the definition~\eqref{eq:shiftradon}, as the delta-function implies that $|w+\vec{x}_0\cdot\unitv{n}|=|\vec{x}\cdot\unitv{n}|\leq1$ for $|\vec{x}|\leq1$. These limits define an integral over a spherical zone between the planes given by $w+\vec{x}_0\cdot\unitv{n}=\pm1$, which are perpendicular to the vector $\vec{x}_0$. The limits are simple to describe in the spherical coordinates if we choose the integration coordinates such that $\vec{x}_0$ is along the $z$-axis, since then they reduce to limits on the $z$-coordinate, such that the integrals can be written as
\begin{equation}
\int_{z_\text{min}}^{z_\text{max}}\int_{0}^{2\pi}P_n(w+x_0z)Y_{lm}(z,\varphi)\difd\varphi\difd z,
\end{equation}
where
\begin{align}
z_\text{min}&=\max\left\{-1,-\frac{1+w}{x_0}\right\},\\
\quad z_\text{max}&=\min\left\{1,\frac{1-w}{x_0}\right\}.
\end{align}
The fact that the azimuthal integral is over $[0,2\pi]$ means that all integrals for $m\neq0$ vanish. What remains is
\begin{equation}
A_{nl}(w,x_0)=2\pi N_l\int_{z_\text{min}}^{z_\text{max}}P_n(w+x_0z)P_l(z)\difd z,
\label{eq:afflegint}
\end{equation}
where $N_l$ is a normalization factor which depends on the spherical harmonic convention used.

Because the coefficients $f'_{nlm}$ were given in an arbitrary coordinate system, use of the expression~\eqref{eq:afflegint} requires them to be rotated to the appropriate integration coordinate system. As discussed in the previous section, the rotation of Zernike functions is effectively just rotation of the spherical harmonic functions, and so can be accomplished via Wigner $D$-matrices. Denoting the rotated coefficients with $f^{(R)}_{nlm}$, we then have
\begin{equation}
\int_{S^2}\mathcal{R}[f](w,\unitv{n})\difd\Omega=2\pi\sum_{n=0}^{L+2}\sum_{l=0}^nf^{(R)}_{nl0}A_{nl}(w,x_0).
\label{eq:radonangle}
\end{equation}

The main challenge in Eq.~\eqref{eq:radonangle} is efficient evaluation of the coefficients $A_{nl}(w,x_0)$. The coefficients $A_{n0}$ and $A_{0l}$ reduce to integrals of Legendre polynomials and may be computed in terms of
\begin{equation}
Q_n(x)=\int_0^xP_n(z)\difd z,
\end{equation}
for which a recursion relation can be derived from the recursion relation of the Legendre polynomials. Using the recursion of Legendre polynomials, it is straightforward to find the five-term recursion
\begin{equation}
A_{nl}=\frac{2n-1}{n}wA_{n-1,l}-\frac{n-1}{n}A_{n-2,l}+\frac{2n-1}{n}\frac{l+1}{2l+1}x_0A_{n-1,l+1}+\frac{2n-1}{n}\frac{l}{2l+1}x_0A_{n-1,l-1}.
\label{eq:afflegrec}
\end{equation}
This recursion works well for the region $w+x_0<1$, where $z_\text{min}=-1$ and $z_\text{max}=1$. However, for $w+x_0>1$, it is highly unstable with an error that grows exponentially with increasing $n$. Consequently, in double precision floating point arithmetic the error exceeds the magnitude of the coefficient after 20 to 80 terms, depending on the parameters $(w,x_0)$. This makes it difficult to apply for even moderately large values of $n$.

Alternative recursive expressions to~\eqref{eq:afflegrec} exist, but we have been unable to find a stable one for every region of the parameter space. Instead, for regions where the recursion is unstable, a hybrid method may be employed. Because the integrand of $A_{nl}$ is a polynomial expression of order $n+l$, Gauss-Legendre quadrature with $(N+1)/2$ nodes gives an exact expression for the integral
\begin{equation}
\int_{z_\text{min}}^{z_\text{max}}P_n(w+x_0z)P_l(z)\difd z=\sum_{i=1}^{\frac{N+1}{2}}w_iP_n(w+x_0z_i)P_l(z_i)
\end{equation}
up to $n+l\leq N$. The values $P_n(w+x_0z_i)$ and $P_l(z_i)$ can be computed recursively for the collections of nodes. However, because direct integration is somewhat expensive for large numbers of nodes, the integrals may be computed for some rows $n$ and $n+1$, and then the recursion~\eqref{eq:afflegrec} may be used for $K$ steps in the forward and backward directions from these rows, with $K$ chosen such that the error from using the recursion remains small.

In the context of dark matter event rate integrals it is common for $(\mathcal{R}f)(w,\unitv{n})$ to be multiplied by some response function $S(w,\unitv{n})$, characterizing the detector's response to the scattering events. The response can originate from a number of material effects. For the purposes of this discussion, it is simply a function of the recoil direction and energy. Multiplying the Radon transform with a response function, the integrand in~\eqref{eq:radonangle} is replaced with
\begin{equation}\label{eq:response_radon_product}
S(w,\unitv{n})\mathcal{R}[f](w,\unitv{n})=2\pi S(w,\unitv{n})\sum_{n=0}^{L+2}\sum_{l=0}^n\sum_{|m|\leq l}f'_{nlm}P_n(w+\vec{x}_0\cdot\unitv{n})Y_{lm}(\unitv{n}).
\end{equation}
The angular integration of this expression proceeds similarly as before, with the added complication that $S(w,\unitv{n})$ also needs to be rotated to the integration coordinates, and then multiplied to the integrand. To accomplish that, we may write $S(w,\unitv{n})$ in terms of its spherical harmonic coefficients $S_{l'm'}(w)$. These coefficients may then be rotated to the integration coordinates, resulting in new coefficients $S^{(R')}_{l'm'}(w)$. The next step is to transform the coefficients $S^{(R')}_{l'm'}(w)$ and $f^{(R)}_{nlm}$ back to the grid on the surface of the sphere, giving the values $S^{(R')}(w,\unitv{n}_{ij})$ and $f^{(R)}_n(\unitv{n}_{ij})$ on the grid points. The grids can then be multiplied together as $f^S_n(w,\unitv{n}_{ij})=S^{(R')}(w,\unitv{n}_{ij})f^{(R)}_n(\unitv{n}_{ij})$. Note that if the spherical expansion of $S(w,\unitv{n})$ has been taken up to $L'$, then in order to avoid aliasing the multiplication requires a grid which allows exact spherical harmonic expansions up to degree $M=L+L'+2$.

In the next step, it would seem intuitive to transform $f^S_n(w,\unitv{n}_{ij})$ back into spherical harmonic coefficients. However, at this point the integrand has the form
\begin{equation}
S(w,\unitv{n})\mathcal{R}[f](w,\unitv{n})=2\pi\sum_{n=0}^{L+2}f^S_n(w,\unitv{n})P_n(w+x_0z),
\end{equation}
where only $f^S_n(w,\unitv{n})$ has dependence on the azimuthal coordinate. Therefore, the azimuthal integration reduces to taking the average over the grid points in the azimuthal direction (zeroth Fourier component),
\begin{equation}
f^S_n(w,z_i)=\frac{1}{2M+1}\sum_{j=1}^{2M+1}f^S_n(w,\unitv{n}_{ij}).
\label{eq:fs-azimuthal-integral}
\end{equation}
The grid $f^S_n(w,z_i)$ can then be used to compute a Legendre polynomial expansion
\begin{equation}
f^S_n(w,z)=\sum_{l=0}^{n+L'}f^S_{nl}(w)P_l(z).
\end{equation}
Note that the expansion only needs to be taken up to $n+L'$. The reason is that $f_n^{(R)}(\hat{n}_{ij})$ has spherical harmonic components only up to degree $n$, and therefore $f^S_n(w,\unitv{n}_{ij})$ only has them up to degree $n+L'$. It then follows that 
\begin{equation}
\int_{S^2}S(w,\unitv{n})\mathcal{R}[f](w,\unitv{n})=2\pi\sum_{n=0}^{L+2}\sum_{l=0}^{n+L'}f^S_{nl}(w)A_{nl}(w,x_0).
\end{equation}

In the nonrelativistic effective theory of DM--nucleus scattering, one also needs to be able to compute the transverse Radon transform~\eqref{eq:transverse-radon} for the shifted distribution, that is the Radon transform of $[(\vec{x}-\vec{x}_0)^2-w^2]f(\vec{x})$. The transverse Radon transform can be expressed as a linear combination of the Radon transforms $\mathcal{R}[f]$, $\mathcal{R}[x_if]$ for $x=1,2,3$, and $\mathcal{R}[x^2f]$. The Zernike functions, again, are advantageous here, because due to their polynomial nature, $x_iZ_{nlm}(\vec{x})$ and $x^2Z_{nlm}(\vec{x})$ can be expressed as finite linear combinations of Zernike functions. Let $f_{nlm}$ be the Zernike coefficients of $f(\vec{x})$, and let $f^{(i)}_{nlm}$ and $f^{(x^2)}_{nlm}$ be the Zernike coefficients of $x_if(\vec{x})$ and $x^2f(\vec{x})$, respectively. It follows that
\begin{equation}
\begin{split}
f^{(i)}_{nl,\pm m}&=A^{(i)}_{---}(n,l,\pm m)f_{n-1,l-1,\pm(m-1)}+A^{(i)}_{--+}(n,l,\pm m)f_{n-1,l-1,\pm(m+1)}\\
&+A^{(i)}_{-+-}(n,l,\pm m)f_{n-1,l+1,\pm(m-1)}+A^{(i)}_{-++}(n,l,\pm m)f_{n-1,l+1,\pm(m+1)}\\
&+A^{(i)}_{+--}(n,l,\pm m)f_{n+1,l-1,\pm(m-1)}+A^{(i)}_{+-+}(n,l,\pm m)f_{n+1,l-1,\pm(m+1)}\\
&+A^{(i)}_{++-}(n,l,\pm m)f_{n+1,l+1,\pm(m-1)}+A^{(i)}_{+++}(n,l,\pm m)f_{n+1,l+1,\pm(m+1)}
\end{split}
\label{eq:transverse-coeff-12}
\end{equation}
for $i=1,2$, where $m\geq0$,
\begin{equation}
\begin{split}
f^{(3)}_{nlm}&=A^{(3)}_{--}(n,l,m)f_{n-1,l-1,m}+A^{(3)}_{-+}(n,l,m)f_{n-1,l+1,m}\\
&+A^{(3)}_{+-}(n,l,m)f_{n+1,l-1,m}+A^{(3)}_{++}(n,l,m)f_{n+1,l+1,m},
\end{split}
\label{eq:transverse-coeff-3}
\end{equation}
and
\begin{equation}
f^{(x^2)}_{nl,\pm m}=A^{(x^2)}_{-}(n,l,m)f_{n-2,lm}+A^{(x^2)}_{0}(n,l,m)f_{nlm}+A^{(x^2)}_{+}(n,l,m)f_{n+2,lm}.
\label{eq:transverse-coeff-x2}
\end{equation}
The coefficients of these relations can be derived from recursion relations of building blocks of the Zernike functions. Their expressions and derivation are found in the Appendix~\ref{app:trradon}.

In summary, the algorithm for evaluating the angle integrated Radon transform---given a distribution $f(\vec{x})$, a response $S(w,\unitv{n})$, a vector $\vec{x}_0$ (corresponding to $\vec{v}_\text{lab}$), and a scalar $w$ (corresponding to $v_\text{min}$)---is as follows
\begin{enumerate}
\item Evaluate the distribution $f(\vec{x})$ on the quadrature grid described in Sec.~\ref{sec:zernike}, and the response $S(w,\unitv{n})$ on a corresponding spherical grid.
\item Compute the Zernike expansion coefficients $f_{nlm}$ from the grid points using Eqs.~\eqref{eq:stepfm}--\eqref{eq:stepfnlm}, and the response coefficients $S_{lm}(w)$ using equations corresponding to~\eqref{eq:stepfm} and~\eqref{eq:stepflm}.
\item Use equation~\eqref{eq:radon-fnlm} to obtain the Radon transform coefficients $f'_{nlm}$ from $f_{nlm}$.
\item Rotate the coefficients $f'_{nlm}$ and $S_{lm}(w)$ to a coordinate system whose $z$-axis is in the direction of $\vec{x}_0$ to obtain the rotated coefficients $f^{(R)}_{nlm}$ and $S^{(R')}_{lm}(w)$.
\item Transform $f^{(R)}_{nlm}$ and $S^{(R')}_{lm}(w)$ back onto the spherical quadrature grid to obtain the values $f^{(R)}_{n}(\unitv{n}_{ij})$ and $S^{(R')}_{lm}(w,\unitv{n}_{ij})$.
\item Multiply $f^{(R)}_{n}(\unitv{n}_{ij})$ and $S^{(R')}_{lm}(w,\unitv{n}_{ij})$ to obtain the products $f^S_n(w,\unitv{n}_{ij})$.
\item Use Eq.~\eqref{eq:fs-azimuthal-integral} to integrate $f^S_n(w,\unitv{n}_{ij})$ over the azimuthal direction to obtain $f^S_n(w,z_i)$.
\item Use the equivalent of Eq.~\eqref{eq:stepflm} to obtain the Legendre polynomial expansion coefficients $f^S_{nl}(w)$ of $f^S_n(w,z_i)$.
\item Use a combination of the recursion relation~\eqref{eq:afflegrec} and Gauss--Legendre quadrature to evaluate the integrals~\eqref{eq:afflegint} to obtain the coefficients $A_{nl}(w,x_0)$.
\item Sum the products $f^S_{nl}(w)A_{nl}(w,x_0)$ over $n$ and $l$ and multiply by $2\pi$.
\end{enumerate}
For the transverse Radon transform, Eqs.~\eqref{eq:transverse-coeff-12}--\eqref{eq:transverse-coeff-x2} need to be used between steps 2 and 3. These coefficients are then combined at appropriate stages. On the other hand, if the response is isotropic, steps related to the response functions can naturally be skipped.

\section{Convergence and performance}
\label{sec:benchmark}

The primary advantage of an expansion of a distribution in a basis of smooth functions is the rapid convergence of the expansion for smooth distributions. With the Zernike basis in particular, the basis function $Z_{nlm}(x,y,z)$ is a polynomial of order $n$ in the Cartesian components. Therefore, a Zernike expansion up to order $L$ is guaranteed to fit a polynomial of the same order exactly. It follows that the expansion truncated to order $L$ for a distribution is as good an approximation as an interpolating polynomial constructed on the grid outlined in Sec.~\ref{sec:zernike}. These properties of the fast convergence of the Zernike expansion for smooth functions also lead to rapid convergence of the Zernike based Radon transform, which is the foundation of the efficiency of this method.

In this section we demonstrate the convergence properties of the approach, as well as the performance of our implementation of the algorithm. All performance benchmarks were executed on an AMD Ryzen 9 7950x CPU and 64 gigabytes of main memory. The code for the benchmarks is available in the repositories of the C$++$ libraries \texttt{ZebraDM} \href{https://github.com/sebsassi/zebradm}{\faGithub}, which implements the Radon transform algorithm, and the companion library \texttt{zest} \href{https://github.com/sebsassi/zest}{\faGithub}, which implements the Zernike and spherical harmonic transforms, as well as the rotation operations needed for implementation of the main algorithm. Significant effort has gone into optimizing the implementations in these libraries, but it is plausible that substantial optimization opportunities remain.

We note that while the 7950x has 16 physical cores, the implementation of the algorithm used for these benchmarks is strictly single-threaded. The problem of computing dark matter rates for parameter scans is of course, in principle, trivially parallelizable. However, for large truncation orders, the algorithm is constrained by memory throughput and size, so in practice there are cases where naive parallelization on a single node may not be viable due to lack of available memory, or may not be useful due to limited memory bandwidth. There are ways of applying a more sophisticated parallelism for the algorithm, which can avoid some of these issues, but we do not pursue parallelization in this work.

To demonstrate these convergence properties and efficiency of the algorithm in practice, we have analyzed the convergence and performance of the algorithm for a collection of test distributions, for which we compute the scattering rate on a two-dimensional grid in time and recoil energy. These test distributions have been chosen to be representative of typical and plausible dark matter velocity distributions, while posing different levels of challenge to numerical integrators. Thus, as distribution one we choose a generic anisotropic Gaussian
\begin{equation}
f_1(\vec{x})=e^{-\frac{1}{2}\vec{x}^T\Sigma^{-1}\vec{x}},
\label{eq:test-f1}
\end{equation}
where the coefficients of the inverse covariance matrix have been set arbitrarily to
\begin{equation}
\Sigma^{-1}=
\begin{pmatrix}
3.0 & 1.4 & 0.5 \\
1.4 & 0.3 & 2.1 \\
0.5 & 2.1 & 1.7 \\
\end{pmatrix}.
\end{equation}
This distribution represents a base case of a distribution with some anisotropic structure.

Distribution two is the collection of four, progressively sharper Gaussians 
\begin{align}
f_2(\vec{x})&=0.4g_1(\vec{x})+0.3g_2(\vec{x})+0.2g_3(\vec{x})+0.1g_4(\vec{x}),\\
\label{eq:test-f2}
g_i(\vec{x})&=\frac{1}{\pi^{3/2}\bar{x}_i^3}\exp\left(\frac{|\vec{x}-\vec{x}_i|^2}{\bar{x}_i^2}\right),
\end{align}
with parameters
\begin{equation}
\begin{gathered}
\vec{x}_1=\frac{1}{537}(0,0,-230),\quad \vec{x}_2=\frac{1}{537}(80,0,-80),\\
\vec{x}_3=\frac{1}{537}(-120,-250,-150),\quad \vec{x}_4=\frac{1}{537}(50,30,-400),
\end{gathered}
\end{equation}
and
\begin{equation}
\bar{x}_1=\frac{220}{537},\quad \bar{x}_2=\frac{70}{537},\quad \bar{x}_3=\frac{50}{537},\quad \bar{x}_4=\frac{25}{537}.
\end{equation}
This distribution was used to test the Haar wavelet method in~\cite{Lillard:2023cyy}. We include it here to provide a point of comparison, and because it gives somewhat of a worst case scenario for the algorithm with its progressively sharpening Gaussians. Note that compared to~\cite{Lillard:2023cyy}, the parameters have been scaled such that $x=1$ corresponds to $v=v_\text{esc}=537$ km/s.

Distribution three is the anisotropic part of the SHM$^{++}$ model~\cite{Evans:2018bqy} as introduced in Eq.~\eqref{eq:anisodist},
\begin{equation}
f_3(\vec{x})=f_A(v_\text{esc}\vec{x}).
\label{eq:test-f3}
\end{equation}
This represents a distribution that appears in actual models of dark matter. It is worth noting that in the real SHM$^{++}$ distribution the anisotropy is weakened by the majority contribution from the isotropic Gaussian term. Therefore, the Zernike expansion of the SHM$^{++}$ converges more rapidly.

In addition, for testing the Zernike-based Radon transform method with anisotropic response functions, we use two artificial test response functions. Due to lack of availability of simple real models of anisotropic responses, these functions have been chosen to demonstrate two boundary cases of a very easy, and a very difficult to integrate response.

The first response is a slowly varying function with a large dipole component,
\begin{equation}
S_1(w,\unitv{n})=\frac{w^4}{1+w^4}e^{2\unitv{n}\cdot\unitv{n}_0},
\end{equation}
where $\unitv{n}_0=\vec{n}_0/|\vec{n}_0|$ is a unit vector with $\vec{n}_0=(1,1,1)$. The second test response is given by
\begin{equation}
S_2(w,\unitv{n})=\Theta_{10}\left(\frac{2}{3}w-W(\unitv{n})\right),
\end{equation}
where
\begin{equation}
W(\unitv{n})=1-e^{2(Y_{6,4}(\unitv{n})-1)},
\end{equation}
and the function
\begin{equation}
\Theta_a(x)=\frac{1}{2}[1 + \tanh(ax)]
\end{equation}
is a smoothed step function, and $Y_{6,4}(\unitv{n})$ is a spherical harmonic. This function has been designed to be difficult to integrate numerically. The smoothed step function in particular ensures there are areas where the function varies rapidly, which numerical integrators struggle with, and which have slowly converging spherical harmonic expansions.
\begin{figure}
\includegraphics{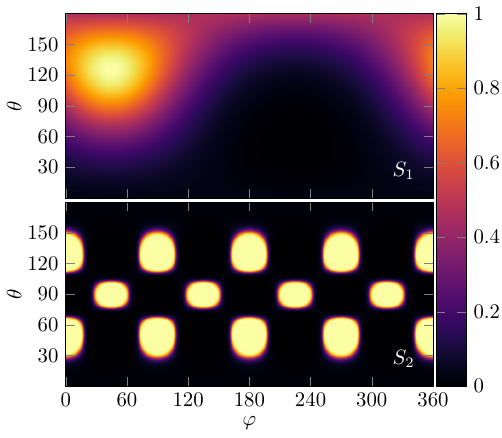}
\caption{Test responses (normalized to maximum value) as a function of the direction $\unitv{n}$ at $w=1$.}
\label{fig:responses}
\end{figure}
The shape of the test responses is shown in Fig.~\ref{fig:responses} for fixed $w=1$.

\begin{figure}
\includegraphics{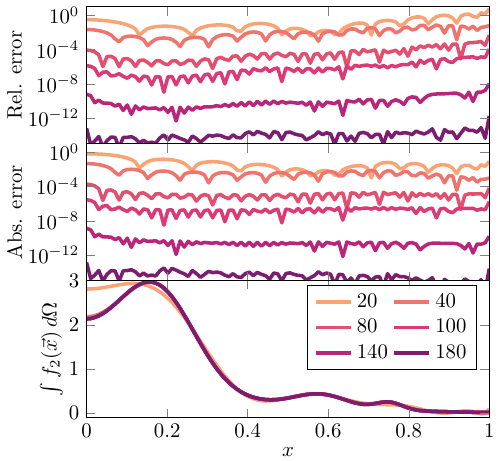}
\caption{Error in the truncated Zernike expansion on the angle integrated test function $F_2(x)=\int f_2(\vec{x})\difd\Omega$ for truncation degrees $L=20,40,80,100,140,180$. From bottom to top: the values $F_2^{(L)}(x)=\sum_{k=0}^{L/2}f^{(2)}_{2k,0,0}R_{2k}^0(x)$, the absolute errors $|F_2(x)-F_2^{(L)}(x)|$, and relative errors $|F_2(x)-F_2^{(L)}(x)|/|F_2(x)|$.}
\label{fig:zernike-radial}
\end{figure}

Of the test distributions presented above, the four-Gaussian distribution $f_2$ is the most challenging numerically due to the presence of the sharp Gaussian $g_4(\vec{x})$. It is therefore a good test case to examine the convergence properties of the Zernike expansion. For visualization purposes, Fig.~\ref{fig:zernike-radial} shows the Zernike approximations $F_2^{(L)}(x)=\sum_{k=0}^{L/2}f^{(2)}_{2k,0,0}R_{2k}^0(x)$ for the angle integrated distribution $F_2(x)=\int f_2(\vec{x})\difd x$, for various truncation values of $L$, as well as the absolute and relative errors of these approximations. One can see that after about degree $L\approx40$, the expansion starts converging exponentially, with roughly an order of magnitude reduction in error for every increase of $L$ by 10, all the way down to a relative error of about $10^{-14}$ at $L=180$.

\begin{figure}
\includegraphics{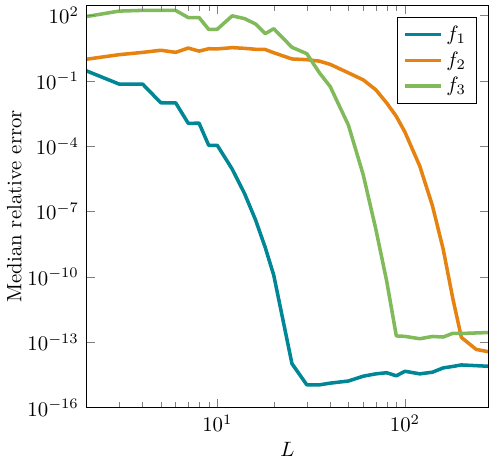}
\caption{Median relative errors of Zernike expansion of the test distributions as a function of the truncation degrees $L$, demonstrating exponential convergence of the Zernike expansion.}
\label{fig:zernike-errors}
\end{figure}

The rapid convergence of the Zernike expansion can be seen more clearly in Fig.~\ref{fig:zernike-errors}, which shows the evolution of the median relative error of the Zernike expansions of the test distributions in terms of the truncation degree $L$. Here it can be seen that the error remains relatively constant until it starts decreasing exponentially, and reaches a minimum due to finite floating point precision. This precision is a couple orders of magnitude above the floating point precision of around $10^{-16}$ due to accumulation of the floating point error in the algorithm. We see that $f_2$ has the most high frequency information, with the error saturating around $L=200$, with $f_1$ and $f_3$ being significantly smoother, reaching saturation around $L=30$ and $L=90$, respectively.

We compare the performance of the Zernike based Radon transform method to the performance of plain numerical integration. To provide a fair comparison between the two approaches, we have also sought to optimize the numerical integration using a custom optimized implementation of adaptive multi-dimensional quadrature using the Genz--Malik integration nodes~\cite{GENZ1980295}. This algorithm uses hyperrectangular integration domains, and therefore none of the integration bounds can depend on the integration variables. Because of this, the integral with an anisotropic response has been implemented as two nested two-dimensional integrals. The isotropic response case, however, can be implemented as a single three-dimensional integral.

The analysis of perfomance is slightly complicated by the fact that the two algorithms work in fundamentally different ways. The basic numerical integration algorithm is very straightforward, as it just sequentially performs the velocity--angle integral for every time--energy pair. Therefore, if we have $N_t$ time points and $N_E$ energy points, and the time to compute a single integral is $T_I$, the total time is simply $T=N_tN_ET_I$.

In contrast, the Zernike-based Radon transform method may be divided into three steps. First, the distribution is evaluated on the quadrature grid needed for the fast Zernike transform. Second is the Zernike transform itself, and last is the evaluation of the Radon transform at a collection of time and energy points. We can therefore express the total computation time as $T=T_G+T_Z+T_R$, where $T_G$ is the grid evaluation time, $T_Z$ is the Zernike transform time, and $T_R$ is the angle integrated Radon transform time. In the event that the parameters of the velocity distribution do not change, only the time $T_R$ has practical relevance. In a general comparison to the basic integration method, however, all need to be accounted for. Furthermore, only $T_G$ depends on the evaluation time of the velocity distribution itself.

Most importantly, however, only $T_R$ depends on $N_E$ and $N_t$. Therefore, asymptotically, for an increasing number of time--energy points, the contribution from $T_R$ is going to dominate. However, this scaling is not simply proportional to the total number of points $N_tN_E$, because some computationally expensive parts of the angle integrated Radon transform---such as the coordinate rotation to match the integration coordinates with direction of $\vec{x}_0$---only need to be performed once for every time point. As a consequence, the algorithm gets more efficient the more energy points there are.

\begin{figure}
\includegraphics{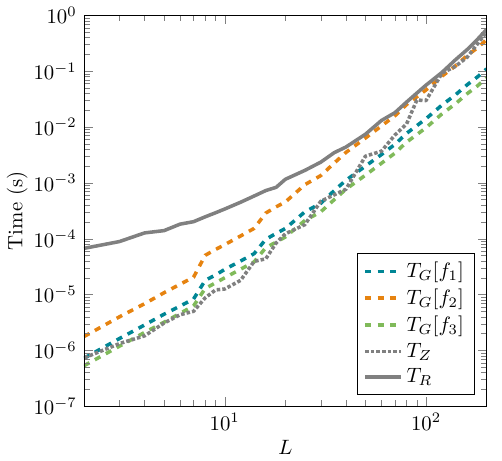}
\caption{Evaluation times of different steps of the Zernike-based Radon transform as a function of the truncation degrees $L$. The time $T_R$ is for an isotropic response with $N_t=10$ and $N_E=100$.}
\label{fig:zernike-times}
\end{figure}

The contributions $T_G$, $T_Z$, and $T_R$ to the total time are shown in Fig.~\ref{fig:zernike-times} in terms of the truncation degree $L$. The times $T_R$ are for an isotropic response with the values $N_t=10$ and $N_E=100$. For an anisotropic response the time $T_R$ also depends on the expansion order of the response, but is always going to be greater than with the isotropic response. The values of $N_t$ and $N_E$ were chosen as representative of a typical order of magnitude one might have to capture the annual modulation of the dark matter event rate. Note, however, that in an analysis studying the daily modulation of dark matter, $N_t$ may be larger by couple orders of magnitude. As can be seen in the figure, $T_R$ typically dominates the total evaluation time, but for large values of $L$ the times $T_Z$ and $T_G$ can make a substantial contribution to the total time. However, as noted above, this is with $N_t=10$ and $N_E=100$. If both of these numbers were doubled, then $T_R$ would constitute about 80\% of the total computation time for all values of $L$ shown here.

In case of an anisotropic response function we have, for completeness, also included time to evaluate the spherical harmonic coefficients $S_{lm}(w)$ in this analysis. However, we note that because the material response typically does not change throughout a direct detection analysis, the spherical harmonic coefficients could typically be precomputed once for the energy grid, and then reused. In a realistic scenario they can therefore be regarded as constant input data. Furthermore, in reality, an anisotropic material response function would likely be in the form of discrete numerical data, which would need to be interpolated or fitted somehow to obtain a smooth function. Therefore, the time it takes to evaluate the coefficients $S_{lm}(w)$ for the test responses is not indicative of the time it would take to do the same for a real response. However, for purposes of comparison, we try to present a scenario that does not unduly disadvantage the conventional numerical integration approach. Therefore we include all the relevant times for the Zernike-based approach.

Because the basic numerical integration approach for evaluating the Radon integrals is so slow, we could not always perform the benchmarks for it with as many time--energy points as we could for the Zernike-based method. For this reason, in order to compare the results, we present the effective integration time defined as
\begin{equation}
T_I^\text{eff}=\frac{T}{N_tN_E}.
\label{Teff}
\end{equation}
The effective integration time is a meaningful metric even if different values of $N_t$ and $N_E$ are used for the two methods, because for the basic numerical integration it is approximately independent of the choice of $N_t$ and $N_E$.

To quantify the performance of the algorithm, we compare the accuracy of the angle integrated Radon transform to the effective integration time. We define the relative error by comparing the results to reference values obtained as follows: First, the velocity distribution is evaluated by the Zernike expansion with $L=200$. As seen in Fig.~\ref{fig:zernike-errors}, at this value the convergence has saturated for each of our test distributions. Second, for the spherical harmonic expansion of the response functions---see the discussion below Eq.~\eqref{eq:response_radon_product}---we choose $L'=800$.

\begin{figure}
\includegraphics{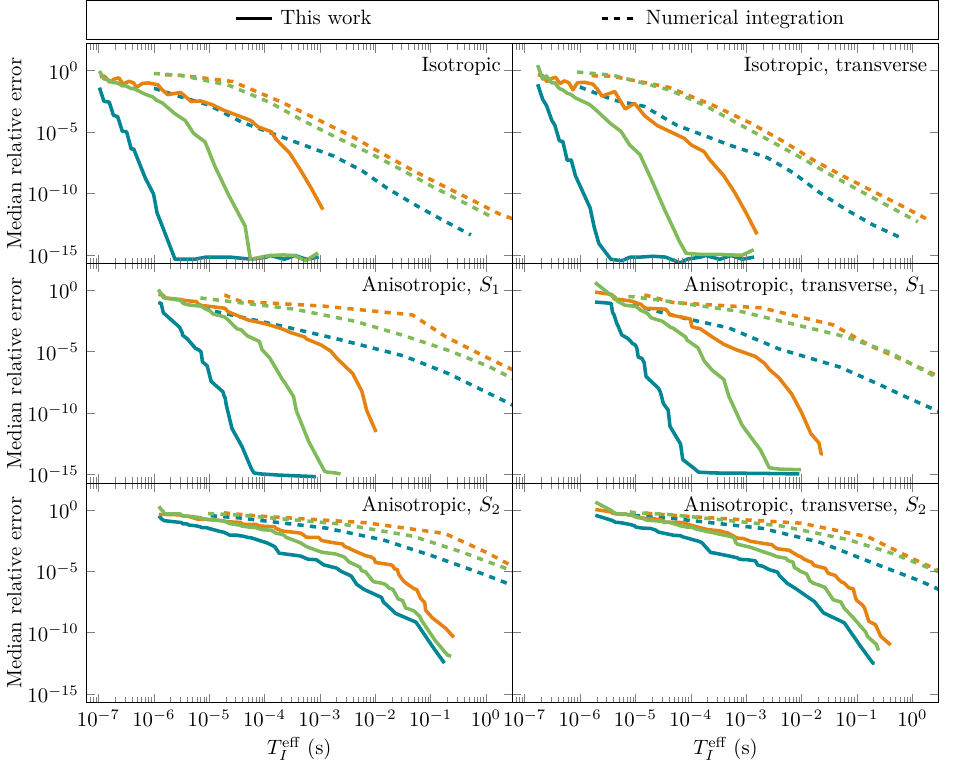}
\caption{Comparison of the performance of the Zernike-based method (this work, solid lines) and direct numerical integration (dashed lines) for the distributions $f_1$, $f_2$, and $f_3$. From top to bottom: isotropic response, response function $S_1$, and response function $S_2$. The left column corresponds to the conventional Radon transform, while the right column corresponds to the transverse Radon transform. The $y$-axis corresponds to the median relative error, while the $x$-axis corresponds to the effective integration time $T_I^\text{eff}$ defined in Eq. (\ref{Teff}).}
\label{fig:zernike-time-comp}
\end{figure}

Fig.~\ref{fig:zernike-time-comp} compares the evolution of the median relative error with the effective integration time $T_I^\text{eff}$ between the Zernike-based method and direct numerical integration. In all cases, for any relative error, the Zernike-based approach is at least an order of magnitude faster than numerical integration. However, due to the rapid, exponential convergence of the approach, it is typically faster by a factor between $10^2$ and $10^4$.

If we consider some notable special cases, for the very smooth anisotropic distribution $f_1$ with an isotropic response, $T_I^\text{eff}$ of around 2 microseconds is sufficient to obtain the minimum error achievable in 64-bit floating point arithmetic with the Zernike-based approach. Generally good enough error of around $10^{-5}$ is reachable within 300 nanoseconds. If we assume $N_tN_E\approx 1000$, at this speed a full year evaluation of the dark matter energy differential rate $dR/dE$ could be done for 3000 points in the parameter space in about 1 second on a single thread. Using conventional numerical integration the same would take around 20 minutes.

Alternatively, if we assume a smooth anisotropic response like $S_1$, a full year daily modulation analysis with $N_t\approx 10^4$ and $N_E\approx20$ for a single mass would take around 1.2 seconds on a single thread with the Zernike-based approach. Doing the same via numerical integration would take around 1 hour and 40 minutes.

When a more difficult response such as $S_2$ is assumed, the Zernike-based approach is naturally slower than in the ideal case of $S_1$, but the relative advantage to basic numerical integration is still retained. It is therefore clear that this method performs much better than conventional numerical integration in all tested cases.

As discussed at the start of this section, the tests above were all performed on a single thread, but it is clear that the problem of evaluating the angle-integrated Radon transform for multiple values is embarrassingly parallel, in principle. In practice, however, memory limitations can come into play. In conventional numerical integration, this is almost never a problem: the amount of memory consumed in an adaptive integration algorithm is proportional to the number of subregions of the integration domain, and since each subregion corresponds to $\mathcal{O}(\text{10--100})$ evaluations of the integrand, and evaluation of the integrand takes some amount of time, the integration is almost always limited by the available time rather than by available memory. Furthermore, evaluating the integrand typically takes a large number of operations compared to the amount of data input and output, cache and memory speed are generally not limiting factors.

In contrast, in the Zernike-based algorithm, a relatively large amount of memory is needed for storing the different expansions and quadrature grids. For example, our implementation of the algorithm necessitates the storage of a spherical quadrature grid with $(M + 1)(2M + 1)$ for each $n\leq L$, where $M=L+L'+2$, with $L$ and $L'$ the truncation degrees of the Zernike expansion of the distribution and spherical harmonic expansion of the response, respectively. For large, though still somewhat realistic, values $L=200$ and $L'=800$ this corresponds to around $4\times 10^8$ values, or around $3.2$ gigabytes of memory. In naive parallelization, this would be duplicated for each thread.

A more significant constraint for the present discussion, arising from this memory usage, however, is that, due to the high computational efficiency of the algorithm, a substantial portion of time is spent on moving data around. As a consequence, when the total amount of required memory exceeds the amount of available cache such that most of the data sits in main memory, the algorithm can eventually become constrained by memory throughput. In this case naive parallelization would be of limited use, since all threads compete for the same memory bandwidth. In fact, this is a factor that can slow down the single-threaded performance of the algorithm for large truncation degrees due to saturation of the memory bus.

This can be regarded as a classic case of memory--computation trade-off: working with transformations on large datasets allows for a drastic reduction in the complexity of the computations. This trade-off evidently pays off as demonstrated by the multiple orders of magnitude improvements over plain numerical integration, but it necessarily leads to a scenario where the limiting factor is data flow.

However, the above limitations only hold for large truncation degrees. In practice, accuracy is typically only needed to a couple of decimal places, and the distributions and responses should generally be fairly smooth. Therefore, it is more typical that $L,L'<100$, or even significantly lower, in which case the required data fits much better into cache, and the speed is not as much restricted by communication with main memory.

\section{Conclusions and outlook}
\label{sec:checkout}

In order to perform parameter scans or likelihood based inference for dark matter direct detection experiments, the event rate needs to be evaluated in a large number of parameter space points. Therefore a fast and efficient evaluation of the event rate is highly desirable.

We have described a novel algorithm for evaluating angle integrated Radon transforms, which can be used in evaluation of dark matter event rates. In particular, we described the algorithm not only for the standard Radon transform, but also for an extension to the so-called transverse Radon transform, which arises in non-relativistic effective theories of dark matter interactions, such that the algorithm is broadly applicable for general effective theory descriptions of dark matter interactions. We demonstrated that, under typical circumstances, this algorithm is faster by a factor between $10^2$ and $10^4$ than the conventional approach using adaptive numerical integration, even if a well-optimized implementation of an advanced numerical integration algorithm is used.

Although the explicit formulas in Sec.~\ref{sec:general} described dark matter scattering primarily in a framework of scattering off nuclear targets, it is worth stressing that the principles of the algorithm we have presented are more generally applicable to any problem, which can be expressed in terms of a three-dimensional Radon transform. Namely, any physical problem where a velocity distribution $f(\vec{v})$ is integrated with a kinematic constraint of the form $\delta(\vec{v}\cdot\unitv{q}-w)$, where $w$ is some term independent of $\vec{v}$. We therefore expect this algorithm to be useful in a broader context than the case of nuclear targets we have used to motivate the algorithm presented in this paper. We expect it to be possible to generalize the algorithm to computation of event rates
for the Migdal effect, DM--electron scattering and DM--phonon scattering. Therefore, the general methods presented in this work can provide a basis of a potential future high-performance framework for general dark matter event rate computations.

We provide an optimized implementation of our algorithm in the C++ library \texttt{ZebraDM}, as well as a separate library \texttt{zest} for working with Zernike and spherical harmonic expansions. Although effort has gone into optimizing these libraries to showcase the potential of this approach, it is plausible that further optimization opportunities remain. In particular, our implementation is single-threaded, while parallelization of some parts of the implementation could be useful when operating on large basis expansions. 
Another possibility which we did not explore in this work is a GPU implementation of the algorithm. It is well-known that algorithm with a high degree of data-parallelism can be up to 10--100 times faster on a good GPU implementation compared to the CPU. Whether or not the approach presented here could significantly benefit from a GPU implementation is a potential avenue of future research.

Another avenue of future research is the choice of basis functions used for the algorithm. Although we strongly argued in favor of the radial Zernike functions in this work, it is plausible that other radial function sets could show other unique benefits, and warrant further study.

\begin{acknowledgments}
This work has been supported by the Research Council of Finland (grant\# 342777). In addition, SS acknowledges the support from the Magnus Ehrnrooth foundation and AA from the Jenny and Antti Wihuri foundation.
\end{acknowledgments}

\pagebreak

\appendix
\section{Transverse Radon transform}
\label{app:trradon}

In the general setting of the nonrelativistic framework of dark matter scattering, the squared scattering amplitude contains terms proportional to $|\vec{v}_\perp|^{2n}$, where $\vec{v}_\perp=\vec{v}-(\vec{v}\cdot\unitv{q})\unitv{q}$. For this reason the general computation of dark matter event rates requires computation of Radon transforms of the form
\begin{equation}
\mathcal{R}_\perp[f](w,\unitv{n})=\int_B\delta(w+\vec{x}_0\cdot\unitv{n}-\vec{x}\cdot\unitv{n})[(\vec{x}-\vec{x}_0)^2-((\vec{x}-\vec{x}_0)\cdot\unitv{n})^2]f(\vec{x})\difd^3x.
\end{equation}
The delta-function forces $(\vec{x}-\vec{x}_0)\cdot\unitv{n}=w$, such that the relevant Radon integral is 
\begin{equation}
\mathcal{R}_\perp[f](w,\unitv{n})=\int_B\delta(w+\vec{x}_0\cdot\unitv{n}-\vec{x}\cdot\unitv{n})[(\vec{x}-\vec{x}_0)^2-w^2]f(\vec{x})\difd^3x.
\end{equation}
Therefore, given the expansion in terms of the Zernike functions, 
\begin{equation}
f(\vec{x})=\sum_{nlm}f_{nlm}Z_{nlm}(\vec{x}),
\end{equation}
we need the expansion of $[(\vec{x}-\vec{x}_0)^2-w^2]f(\vec{x})$. Expansion of the prefactor $(\vec{x}-\vec{x}_0)^2-w^2$ shows that this term can be expressed as a linear combination of the functions $f(\vec{x})$, $x_if(\vec{x})$, and $|\vec{x}|^2f(\vec{x})$, where $x_i$ are the components of $\vec{x}$. Therefore, the problem reduces to the computation of the Zernike expansions of these terms.

The Zernike functions are written in terms of their radial, azimuthal, and polar components as
\begin{equation}
Z_{nlm}(\vec{x})=R_n^l(\rho)\bar{P}_l^{|m|}(\cos\theta)
\begin{cases}
\cos m\varphi,\quad m\geq0\\
\sin |m|\varphi,\quad m<0,
\end{cases}
\end{equation}
where $\bar{P}_l^{|m|}(t)$ are the associated Legendre polynomials, normalized such that
\begin{equation}
\bar{P}_l^{|m|}(t)=
\begin{cases}
\sqrt{2(2l+1)\frac{(l-m)!}{(l+m)!}}P_l^m(t), |m|>0,\\
\sqrt{(2l+1)\frac{(l-m)!}{(l+m)!}}P_l^m(t), m=0,\\
\end{cases}
\end{equation}
where $P_l^m(t)$ are the unnormalized associated Legendre polynomials.

The Cartesian coordinates $x_i$ can be expanded in spherical coordinates. In doing so, besides the radial argument $\rho=|\vec{x}|$, it is useful to write them in terms of the variables $t=\cos\theta$ and $u=\sin\theta=(1-t^2)^{1/2}$, such that
\begin{equation}
x=\rho u\cos\varphi,\quad y=\rho u\sin\varphi,\quad z=\rho t.
\end{equation}
The problem is then reduced to finding expressions as linear combinations of Zernike functions for
\begin{equation}
\rho^2Z_{nlm}(\rho,t\varphi),\quad \rho u\cos\varphi Z_{nlm}(\rho,t\varphi),\quad \rho u\sin\varphi Z_{nlm}(\rho,t\varphi),\quad \rho tZ_{nlm}(\rho,t\varphi).
\end{equation}
This can be done using relevant recursion relations for the functions $R_n^l(\rho)$, $\bar{P}_l^{m}(t)$, $\cos m\varphi$, and $\sin m\varphi$.

For the first one, $\rho^2Z_{nlm}(\rho,t\varphi)$, we can rearrange Eq.~\eqref{eq:radialzernike} to write
\begin{equation}
\rho^2R_n^l(\rho)=J_1^{nl}R_{n+2}^l(\rho)+J_2^{nl}R_n^l(\rho)+J_3^{nl}R_{n-2}^l(\rho),
\end{equation}
where
\begin{align}
J_1^{nl}&=\frac{(n-l+2)(n+l+3)}{(2n+3)(2n+5)},\\
J_2^{nl}&=\frac{(2l+1)^2+(2n+1)(2n+5)}{2(2n+1)(2n+5)},\\
J_3^{nl}&=\frac{(n-l)(n+l+1)}{(2n+1)(2n+3)}.
\end{align}
This relation, however, is only valid for $l\geq n-4$. For the remaining two cases we have
\begin{align}
R_n^n(\rho)&=r^2R_{n-2}^{n-2}(\rho),\\
R_n^{n-2}(\rho)&=((n+\tfrac{1}{2})\rho^2-(n-\tfrac{1}{2}))R_{n-2}^{n-2}(\rho).
\end{align}

The terms which are linear in the Cartesian coordinates require recursion relations for all sets of functions. For the trigonometric functions it is straightforward to use the well-known relations
\begin{align}
\cos\varphi\cos(m\varphi)&=+\tfrac{1}{2}[\cos((m+1)\varphi)+\cos((m-1)\varphi)],\\
\cos\varphi\sin(m\varphi)&=+\tfrac{1}{2}[\sin((m+1)\varphi)+\sin((m-1)\varphi)],\\
\sin\varphi\cos(m\varphi)&=+\tfrac{1}{2}[\sin((m+1)\varphi)-\sin((m-1)\varphi)],\\
\sin\varphi\sin(m\varphi)&=-\tfrac{1}{2}[\cos((m+1)\varphi)-\cos((m-1)\varphi)].
\end{align}
For the associated Legendre polynomials it is possible to manipulate their recursion relations to write
\begin{align}
u\bar{P}_l^m(t)&=k_+(m)\left[\sqrt{\frac{(l-m-1)(l-m)}{(2l-1)(2l+1)}}\bar{P}_{l-1}^{m+1}(t)-\sqrt{\frac{(l+m+1)(l+m+2)}{(2l+1)(2l+3)}}\bar{P}_{l+1}^{m+1}(t)\right],\\
u\bar{P}_l^m(t)&=k_-(m)\left[\sqrt{\frac{(l-m+1)(l-m+2)}{(2l+1)(2l+3)}}\bar{P}_{l+1}^{m-1}(t)-\sqrt{\frac{(l+m-1)(l+m)}{(2l-1)(2l+1)}}\bar{P}_{l-1}^{m-1}(t)\right],\\
t\bar{P}_l^m(t)&=\sqrt{\frac{(l-m+1)(l+m+1)}{(2l+1)(2l+3)}}\bar{P}_{l+1}^m(t)+\sqrt{\frac{(l-m)(l+m)}{(2l-1)(2l+1)}}\bar{P}_{l-1}^m(t),
\end{align}
where $k_+(0)=1/\sqrt{2}$ and $k_+(m>0)=1$ otherwise and $k_-(1)=\sqrt{2}$ and $k_-(m>1)=1$. Finally, for the Zernike functions, which are defined in terms of the Jacobi polynomials in Eq.~\eqref{eq:radialzernike}, we can use the two recursion relations
\begin{align}
(2k+\alpha+\beta+1)P_k^{(\alpha,\beta)}(x)&=(k+\alpha+\beta+1)P_k^{(\alpha,\beta+1)}(x)+(k+\alpha)P_{k-1}^{(\alpha,\beta+1)}(x),\\
(2k+\alpha+\beta+1)\tfrac{1}{2}(1+x)P_k^{(\alpha,\beta)}(x)&=(k+1)P_{k+1}^{(\alpha,\beta-1)}(x)+(k+\beta)P_k^{(\alpha,\beta-1)}(x).
\end{align}
The Zernike functions correspond to $\alpha=0$, $\beta=l+1/2$, $k=(n-l)/2$, and $x=2\rho^2-1$. Plugging these in and multiplying by the appropriate power of $\rho$ gives
\begin{align}
\rho R_n^l(\rho)&=\frac{n+l+3}{2n+3}R_{n+1}^{l+1}(\rho)+\frac{n-l}{2n+3}R_{n-1}^{l+1}(\rho),\\
\rho R_n^l(\rho)&=\frac{n-l+2}{2n+3}R_{n+1}^{l-1}(\rho)+\frac{n+l+1}{2n+3}R_{n-1}^{l-1}(\rho).
\end{align}

Combining the above recursion relations, it is possible to find recursion relations which allow us to write $\rho^2Z_{nlm}(\rho,t\varphi)$ and $x_iZ_{nlm}(\rho,t\varphi)$ as linear combinations of Zernike functions. To conserve space, we do not write down these recursion relations explicitly here. Instead, we will give the coefficients $A^{(1,2)}_{\pm\pm\pm}$, $A^{(3)}_{\pm\pm}$, $A^{(x^2)}_{\pm,0}$ of Eqs.~\eqref{eq:transverse-coeff-12}--\eqref{eq:transverse-coeff-x2}, which express coefficients of the expansions
\begin{align}
\rho^2f(\vec{x})&=\sum_{nlm}f^{(\rho^2)}_{nlm}Z_{nlm}(\vec{x}),\\
x_if(\vec{x})&=\sum_{nlm}f^{(i)}_{nlm}Z_{nlm}(\vec{x}),
\end{align}
in terms of $f_{nlm}$. Thus, for the linear components of $A^{(1)}$ we have
\begin{align}
A^{(1)}_{---}(n,l,\pm m)&=-g_+(m-1)\sqrt{\frac{(l+m-1)(l+m)(n+l+1)^2}{(2n+1)(2n+3)(2l-1)(2l+1)}},\\
A^{(1)}_{--+}(n,l,\pm m)&=+g_-(m+1)\sqrt{\frac{(l-m-1)(l-m)(n+l+1)^2}{(2n+1)(2n+3)(2l-1)(2l+1)}},\\
A^{(1)}_{-+-}(n,l,\pm m)&=+g_+(m-1)\sqrt{\frac{(l-m+1)(l-m+2)(n-l)^2}{(2n+1)(2n+3)(2l+1)(2l+3)}},\\
A^{(1)}_{-++}(n,l,\pm m)&=-g_-(m+1)\sqrt{\frac{(l+m+1)(l+m+2)(n-l)^2}{(2n+1)(2n+3)(2l+1)(2l+3)}},\\
A^{(1)}_{+--}(n,l,\pm m)&=-g_+(m-1)\sqrt{\frac{(l+m-1)(l+m)(n-l+2)^2}{(2n+3)(2n+5)(2l-1)(2l+1)}},\\
A^{(1)}_{+-+}(n,l,\pm m)&=+g_-(m+1)\sqrt{\frac{(l-m-1)(l-m)(n-l+2)^2}{(2n+3)(2n+5)(2l-1)(2l+1)}},\\
A^{(1)}_{++-}(n,l,\pm m)&=+g_+(m-1)\sqrt{\frac{(l-m+1)(l-m+2)(n+l+3)^2}{(2n+3)(2n+5)(2l+1)(2l+3)}},\\
A^{(1)}_{+++}(n,l,\pm m)&=-g_-(m+1)\sqrt{\frac{(l+m+1)(l+m+2)(n+l+3)^2}{(2n+3)(2n+5)(2l+1)(2l+3)}}.
\end{align}
Similarly, for the components of $A^{(2)}$
\begin{align}
A^{(2)}_{---}(n,l,\pm m)&=\pm g_+(m-1)\sqrt{\frac{(l+m-1)(l+m)(n+l+1)^2}{(2n+1)(2n+3)(2l-1)(2l+1)}},\\
A^{(2)}_{--+}(n,l,\pm m)&=\pm g_-(m+1)\sqrt{\frac{(l-m-1)(l-m)(n+l+1)^2}{(2n+1)(2n+3)(2l-1)(2l+1)}},\\
A^{(2)}_{-+-}(n,l,\pm m)&=\mp g_+(m-1)\sqrt{\frac{(l-m+1)(l-m+2)(n-l)^2}{(2n+1)(2n+3)(2l+1)(2l+3)}},\\
A^{(2)}_{-++}(n,l,\pm m)&=\mp g_-(m+1)\sqrt{\frac{(l+m+1)(l+m+2)(n-l)^2}{(2n+1)(2n+3)(2l+1)(2l+3)}},\\
A^{(2)}_{+--}(n,l,\pm m)&=\pm g_+(m-1)\sqrt{\frac{(l+m-1)(l+m)(n-l+2)^2}{(2n+3)(2n+5)(2l-1)(2l+1)}},\\
A^{(2)}_{+-+}(n,l,\pm m)&=\pm g_-(m+1)\sqrt{\frac{(l-m-1)(l-m)(n-l+2)^2}{(2n+3)(2n+5)(2l-1)(2l+1)}},\\
A^{(2)}_{++-}(n,l,\pm m)&=\mp g_+(m-1)\sqrt{\frac{(l-m+1)(l-m+2)(n+l+3)^2}{(2n+3)(2n+5)(2l+1)(2l+3)}},\\
A^{(2)}_{+++}(n,l,\pm m)&=\mp g_-(m+1)\sqrt{\frac{(l+m+1)(l+m+2)(n+l+3)^2}{(2n+3)(2n+5)(2l+1)(2l+3)}}.
\end{align}
Finally, the components of $A^{(3)}$ are
\begin{align}
A^{(3)}_{--}(n,l,\pm m)&=\sqrt{\frac{(l-m)(l+m)(n+l+1)^2}{(2n+1)(2n+3)(2l-1)(2l+1)}},\\
A^{(3)}_{+-}(n,l,\pm m)&=\sqrt{\frac{(l-m)(l+m)(n-l+2)^2}{(2n+3)(2n+5)(2l-1)(2l+1)}},\\
A^{(3)}_{-+}(n,l,\pm m)&=\sqrt{\frac{(l-m+1)(l+m+1)(n-l)^2}{(2n+1)(2n+3)(2l+1)(2l+3)}},\\
A^{(3)}_{++}(n,l,\pm m)&=\sqrt{\frac{(l-m+1)(l+m+1)(n+l+3)^2}{(2n+3)(2n+5)(2l+1)(2l+3)}}.
\end{align}
Here $m\geq0$, and $g_+(0)=g_-(1)=1/\sqrt{2}$ and $g_+(m>0)=g_-(m>1)=1/2$.

For the quadratic component we have
\begin{align}
A^{(\rho^2)}_-(n,l,m)&=\frac{(n-l)(n+l+1)}{\sqrt{(2n-1)(2n+1)^2(2n+3)}},\\
A^{(\rho^2)}_0(n,l,m)&=\frac{(2l+1)^2+(2n+1)(2n+5)}{2(2n+1)(2n+5)},\\
A^{(\rho^2)}_+(n,l,m)&=\frac{(n-l+2)(n+l+3)}{\sqrt{(2n+3)(2n+5)^2(2n+7)}},
\end{align}
valid for $l\leq n-4$, with
\begin{align}
A^{(\rho^2)}_-(n,n-2,m)&=\frac{2}{2n+1}\sqrt{\frac{2n-1}{2n+3}},\\
A^{(\rho^2)}_0(n,n-2,m)&=\frac{(2l+1)^2+(2n+1)(2n+5)}{2(2n+1)(2n+5)},\\
A^{(\rho^2)}_+(n,n-2,m)&=\frac{(n-l+2)(n+l+3)}{\sqrt{(2n+3)(2n+5)^2(2n+7)}},
\end{align}
and
\begin{align}
A^{(\rho^2)}_-(n,n,m)&=0,\\
A^{(\rho^2)}_0(n,n,m)&=\frac{2n+3}{2n+5},\\
A^{(\rho^2)}_+(n,n,m)&=\frac{(n-l+2)(n+l+3)}{\sqrt{(2n+3)(2n+5)^2(2n+7)}}.
\end{align}


\bibliography{bibliography.bib}

\end{document}